\documentclass[a4paper,onecolumn,11pt,accepted=2022-01-21]{quantumarticle}
\pdfoutput=1
\usepackage[utf8]{inputenc}
\usepackage{graphicx}
\usepackage{amssymb,amsmath,amsthm,amsfonts}
\usepackage{hyperref}
\usepackage[numbers,sort&compress]{natbib}
\usepackage[capitalise,compress]{cleveref}
\usepackage{bm}
\usepackage{float}
\usepackage{color}
\usepackage{tikz}
\usetikzlibrary{arrows.meta, bending, matrix, positioning}

\usepackage{tikz}
\usepackage{quantikz}

\begin{document}
\title{Efficient quantum programming using EASE gates on a trapped-ion quantum computer}
\author{Nikodem Grzesiak}
\email{nikodem.grzesiak@gmail.com} 
\affiliation{IonQ, College Park, MD 20740, USA}
\orcid{0000-0002-9016-2673}

\author{Andrii Maksymov}
\email{andrii.maksymov@gmail.com}
\affiliation{IonQ, College Park, MD 20740, USA}
\orcid{0000-0002-8244-2851}

\author{Pradeep Niroula}
\email{pniroula@umd.edu}
\affiliation{Joint Center for Quantum Information and Computer Science, NIST/University of Maryland, College Park, MD 20742, USA}
\affiliation{Joint Quantum Institute, University of Maryland, College Park, MD 20742, USA}
\orcid{0000-0001-8941-7774}

\author{Yunseong Nam}
\email{ynam@umd.edu}
\affiliation{IonQ, College Park, MD 20740, USA}
\affiliation{Department of Physics,
University of Maryland, College Park, MD 20742, USA}
\orcid{0000-0002-2742-3447}

\maketitle

\begin{abstract}
Parallel operations in conventional computing have proven to be an essential tool for efficient and practical computation, and the story is not different for quantum computing. Indeed, there exists a large body of works that study advantages of parallel implementations of quantum gates for efficient quantum circuit implementations. Here, we focus on the recently invented efficient, arbitrary, simultaneously entangling (EASE) gates, available on a trapped-ion quantum computer. Leveraging its flexibility in selecting arbitrary pairs of qubits to be coupled with any degrees of entanglement, all in parallel, we show an $n$-qubit Clifford circuit can be implemented using $6\log(n)$ EASE gates, an $n$-qubit multiply-controlled NOT gate can be implemented using $3n/2$ EASE gates, and an $n$-qubit permutation can be implemented using six EASE gates. We discuss their implications to near-term quantum chemistry simulations and the state of the art pattern matching algorithm. Given Clifford + multiply-controlled NOT gates form a universal gate set for quantum computing, our results imply efficient quantum computation by EASE gates, in general.
\end{abstract}

\section{Introduction}

A gate-based, universal quantum computer is increasingly becoming a commodity between researchers in a variety of fields, and in all instances of use cases, the process of using a quantum computer inevitably includes the key step of compiling a program to an executable targeted to backend quantum hardware. With different architectures available today, leveraging unique capabilities that each different architecture offers is a step to be considered seriously, should one wish to harness the most out of a quantum computer. A parallel may be drawn to conventional computing, where, for example, single instruction multiple data (SIMD) techniques have been used to perform efficient computation \cite{hughes2015single}.

In this paper, we consider a trapped-ion quantum computer that offers a powerful SIMD-like technique called efficient, arbitrary, simultaneously entangling (EASE) gate, as its native operation. As detailed in Sec.~\ref{sec:EASE} (see also \cite{EASE}), an EASE gate can entangle arbitrarily selected pairs of qubits in one step. When the entanglement couplings for all pairs are identical, the EASE gate becomes the so-called global M{\o}lmer-S{\o}rensen (GMS) gate, investigated in \cite{GMS1,GMS2,GMS3} to demonstrate more efficient compilation of quantum programs when compared to a serial approach.

Specific to this work, we focus on three important classes of circuits, widely used in many quantum programs: Clifford circuits, multiply-controlled NOT (NOT$:=\scriptsize{\left(\begin{matrix} 0 \, &1\, \\ 1 \, &0 \, \end{matrix}\right)}$) operation, and the permutation operator. For the first, we consider the number of EASE gates sufficient to implement an $n$-qubit arbitrary Clifford circuit. Specifically, in Sec.~\ref{sec:Clifford}, we improve the best-known bound of $6n$, reported in \cite{GMS3}, to $6\log(n)$, where we dropped additive constant dependence. For the second, we consider an $n$-controlled NOT gate and show, in Sec.~\ref{sec:Toffoli}, the number of EASE gates required can be reduced to $3n/2$. This may be compared to the state-of-the-art $2n$~\cite{GMS2} for the GMS-based approach. Further, we show the number of EASE gates can be as small as two for an arbitrarily large $n$, should the ancillae be inexpensive. Lastly, in Sec.~\ref{sec:permutation} we show a permutation of arbitrary number of qubits can be performed using six EASE gates only.

\section{EASE gates}
\label{sec:EASE}

We start by defining an EASE gate, i.e.,
\begin{equation}
{\rm EASE}\left(\vec{\phi},\vec{\theta}\right) := \prod_{j>k} \exp\left( -i\sigma_{\phi_j}^{(j)}\sigma_{\phi_k}^{(k)} \theta_{jk} / 2 \right),
\end{equation}
where 
\begin{equation}
\sigma_{\phi_j}^{(j)} = \cos(\phi_j) \sigma_x^{(j)} + \sin(\phi_j) \sigma_y^{(j)}
\end{equation}
is a Pauli operator, defined over a vector that points to the equator of a Bloch sphere with azimuthal angle $\phi_j$, acting on qubit $j$ and free parameters $\theta_{jk}$ are the entanglement coupling between qubits $j$ and $k$. Shown in \cite{EASE} was that, even though the number of $\theta_{jk}$ parameters increases quadratically in the number of qubits $n$, the complexity of the control signal design scales at most linearly in the number of qubits $n$, bounded from above by $3n-1$. Note a single two-qubit gate requires similar complexity, i.e., $2n+1$~\cite{Zhu_2006}. In other words, an EASE gate's complexity is at most a small ($<1.5$) multiplicative constant factor more than that of the conventional two-qubit gate. 

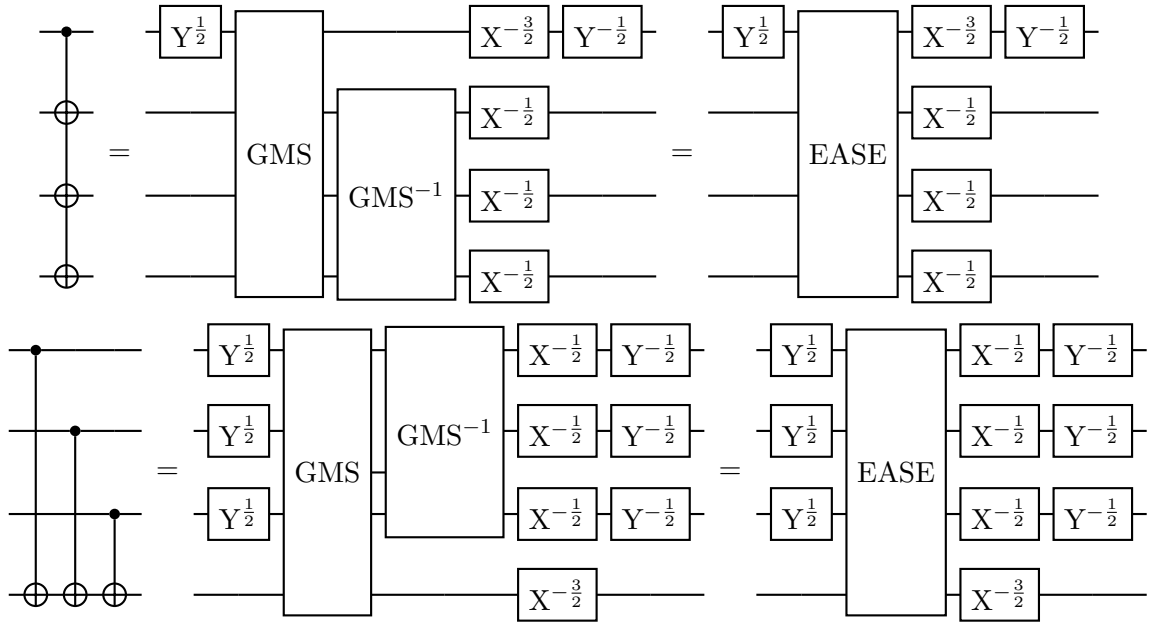
\begin{figure*}
    \centering
    \begin{quantikz}[row sep=1.0em, column sep=0.5em]
    & \ctrl{1} & \qw & & & \gate{\text{Y}^\frac{1}{2}} & \gate[5,nwires={3}]{\text{GMS}} & \qw & \gate{\text{X}^{-\frac{3}{2}}} & \gate{\text{Y}^{-\frac{1}{2}}} & \qw & & & \gate{\text{Y}^\frac{1}{2}}  & \gate[5,nwires={3}]{\text{EASE}} & \gate{\text{X}^{-\frac{3}{2}}} & \gate{\text{Y}^{-\frac{1}{2}}} & \qw \\
    &  \targ{} \vqw{2} & \qw & & & \qw & & \gate[4,nwires={2}]{\text{GMS}^{-1}} & \gate{\text{X}^{-\frac{1}{2}}} & \qw & \qw & & & \qw &  & \gate{\text{X}^{-\frac{1}{2}}} & \qw & \qw \\[-0.25cm]
    & & & = & & & & & & & & = & & &  & & & \\[-0.25cm]
    &  \targ{} \vqw{1} & \qw & & & \qw & & & \gate{\text{X}^{-\frac{1}{2}}} & \qw & \qw & & & \qw &  & \gate{\text{X}^{-\frac{1}{2}}} & \qw & \qw \\
    &  \targ{} & \qw & & & \qw & & & \gate{\text{X}^{-\frac{1}{2}}} & \qw & \qw & & & \qw &  & \gate{\text{X}^{-\frac{1}{2}}} & \qw & \qw
    \end{quantikz}
    
    \begin{quantikz}[row sep=1.0em, column sep=0.5em]
    & \ctrl{4} & \qw & \qw & \qw & & & \gate{\text{Y}^\frac{1}{2}} & \gate[5,nwires={3}]{\text{GMS}} & \gate[4,nwires={2}]{\text{GMS}^{-1}}  & \gate{\text{X}^{-\frac{1}{2}}} & \gate{\text{Y}^{-\frac{1}{2}}} & \qw & & & \gate{\text{Y}^\frac{1}{2}}  & \gate[5,nwires={3}]{\text{EASE}} & \gate{\text{X}^{-\frac{1}{2}}} & \gate{\text{Y}^{-\frac{1}{2}}} & \qw \\
    & \qw & \ctrl{3} & \qw & \qw & & & \gate{\text{Y}^\frac{1}{2}} & & & \gate{\text{X}^{-\frac{1}{2}}} & \gate{\text{Y}^{-\frac{1}{2}}} & \qw & & & \gate{\text{Y}^\frac{1}{2}} &  & \gate{\text{X}^{-\frac{1}{2}}} & \gate{\text{Y}^{-\frac{1}{2}}} & \qw \\[-0.25cm]
    & & & & & = & & & & & & & & = & & &  & & & \\[-0.25cm]
    &  \qw & \qw & \ctrl{1} & \qw & & & \gate{\text{Y}^\frac{1}{2}} & & & \gate{\text{X}^{-\frac{1}{2}}} & \gate{\text{Y}^{-\frac{1}{2}}} & \qw & & & \gate{\text{Y}^\frac{1}{2}} &  & \gate{\text{X}^{-\frac{1}{2}}} & \gate{\text{Y}^{-\frac{1}{2}}} & \qw \\
    &  \targ{} & \targ{} & \targ{} & \qw & & & \qw & & \qw & \gate{\text{X}^{-\frac{3}{2}}} & \qw & \qw & & & \qw &  & \gate{\text{X}^{-\frac{3}{2}}} & \qw & \qw 
    \end{quantikz}
    \caption{Fan-out and fan-in operation implementation using GMS vs. EASE gates for a four-qubit instance. Both GMS and EASE gates used here assume the entangling angle of $\pi/2$ between pairs of qubits. ${\mathrm X}^c$ and ${\mathrm Y}^c$ denote the single-qubit operators $\exp(-i\sigma_x c \pi /2)$ and $\exp(-i\sigma_y c \pi /2)$, respectively. The EASE gate used for the fan-out simultaneously entangles only qubits $(0,1)$, $(0,2)$, and $(0,3)$, where qubits are labeled from zero to three from the top. The EASE gate used for the fan-in simultaneously entangles only qubits $(0,3)$, $(1,3)$, and $(2,3)$.}
    \label{fig:fanout}
    \vspace{-1em}
\end{figure*}

It thus stands to reason that the use of EASE gates, whenever possible, should be explored.\footnote{We note that EASE gates could, e.g., be of lower fidelity than the conventional two-qubit gate in today's trapped-ion quantum computers, as observed in \cite{EASE}. Therefore, at least for the near term, a care needs to taken when determining if the use of EASE gates would indeed be beneficial.} The main differences between the EASE gates and the GMS gates are that (i) one can choose arbitrary pairs to be entangled in an EASE gate whereas one must entangle all pairs in the GMS gate\footnote{We abuse the notion GMS, including the case where only a subset of qubits is considered.} and (ii) the entanglement couplings can be chosen flexibly for each selected pair in an EASE gate whereas in the GMS gate they cannot be. In other words, a GMS gate is a special EASE gate with $\theta_{jk}$ for all $j$ and $k$ being identical to one another. We note that, in this paper, we judiciously use both (i) and (ii) to improve the quantum circuit efficiency.

Due to their versatile use throughout this paper, we briefly discuss the complexity of implementing a fan-in or a fan-out controlled-NOT (CNOT) operation in the following. It was reported in \cite{GMS1} that two GMS gates are needed to implement either the fan-in or the fan-out operation over $n$ qubits. Note a serial approach would require $n-1$ XX($\pi/2$) gates each, where an XX($\theta$) gate is defined as $\exp\left(-i\sigma_{x}\sigma_{x} \theta / 2 \right)$. Leveraging the ability to target arbitrary pairs of qubits, only a single EASE gate is required for each. See Fig.~\ref{fig:fanout} for details. Briefly, an EASE gate can implement any combinations of XX gates in parallel, colliding or not at a qubit. All of the XX gates used to implement either the fan-in or the fan-out operation would thus be implemented using a single EASE gate. Parallel fan-ins to different targets with potentially colliding controls or parallel fan-outs from different controls to potentially colliding targets would also cost only a single EASE gate.

\section{Clifford circuits}
\label{sec:Clifford}

We report in this section an EASE-based method to synthesize an arbitrary Clifford circuit, using a normal form of H-S-CZ-CNOT-H-CZ-S-H~\cite{CliffordNormalForm}. Here, H and S denote single-qubit gate layers of Hadamard ${\mathrm H} := \frac{1}{\sqrt{2}} \scriptsize{\left(\begin{matrix} 1 \, &1\, \\ 1 \, &-1 \, \end{matrix}\right)}$ and phase gates ${\mathrm S} := \scriptsize{\left(\begin{matrix} 1 \, &0\, \\ 0 \, &i \, \end{matrix}\right)}$, respectively, which are trivially parallelizable. Thus, we focus in the following an efficient method to implement CZ -- which stands for controlled-Z, where Z$:=\scriptsize{\left(\begin{matrix} 1 \, &0\, \\ 0 \, &-1 \, \end{matrix}\right)}$, -- and CNOT layers.

\begin{figure}
    \centering
    \begin{quantikz}[row sep=1.0em, column sep=0.5em]
& \ctrl{2} & \qw & & & \gate[3,nwires={2}]{\text{ZZ}} & \gate{\text{S}^{-1}} & \qw & & & \gate{\text{H}} & \gate[3,nwires={2}]{\text{XX}} & \gate{\text{H}}& \gate{\text{S}^{-1}}  &  \qw\\[-0.25cm]
& & & = & & & & & = & & & & & & \\[-0.25cm]
& \gate{\text{Z}} & \qw & & & & \gate{\text{S}^{-1}} & \qw & & &\gate{\text{H}} & & \gate{\text{H}} & \gate{\text{S}^{-1}} & \qw
    \end{quantikz}
\caption{\label{CZdecomp} Decomposition of a CZ gate using ZZ and ${\mathrm S}^{-1}$ gates, and then into XX, H, and ${\mathrm S}^{-1}$ gates. XX and ZZ gates here assume an entanglement angle $\theta = \pi/2$.}
\end{figure}
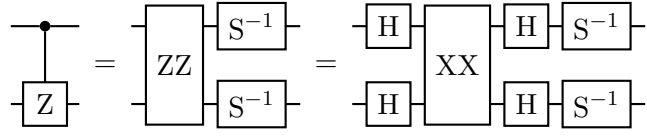

A CZ layer is straightforward to implement using a single EASE gate. This is so since, as shown in Fig.~\ref{CZdecomp}, a CZ gate can be decomposed into a ZZ$:=\exp\left(-i\sigma_{z}\sigma_{z} \theta / 2 \right)$ gate and two ${\mathrm S}^{-1}$ gates, where ZZ and ${\mathrm S}^{-1}$ gates commute with one another. This way, all of the ZZ gates can be layered to form a single block of ZZ gates and the aggregated ${\mathrm S}^{-1}$ gates, forming a single-qubit gate layer, can be implemented straightforwardly in parallel. The single block of ZZ gates is nothing but a Hadamard-layer conjugated XX gates, and our EASE gates can implement multiple XX gates over arbitrary pairs in a single step.

We next turn to an efficient implementation of a CNOT layer. Discussed in \cite{GMS1} was the implementation detail of a CNOT layer, where the layer was LU-decomposed first, then each of the two resulting ``triangular'' transformations of size $n \times n$ was implemented using $2n$ GMS gates, where two GMS gates were spent for each of the $n$ fan-out CNOTs. This results in $4n$ GMS gates required for a CNOT layer. A naive replacement of two GMS gates used in each fanout with one EASE, possible due to the aforementioned flexibility of the EASE to apply XX gates only between the qubit pairs that comprise of the shared control and all of the targets, would result in $2n$ EASE gates per CNOT layer.

\begin{figure*}
\[
\begin{tikzpicture}[
    node distance=1mm and 0mm,
    baseline]
\matrix (M1) [matrix of nodes,{left delimiter=[},{right delimiter=]}]
{
    1 & 0 & 1 & 1 & 1 & 0 & 1 & 1 \\ 
      & 1 & 1 & 0 & 0 & 1 & 0 & 1 \\ 
      &   & 1 & 1 & 1 & 0 & 1 & 1 \\ 
      &   &   & 1 & 1 & 1 & 1 & 1 \\ 
      &   &   &   & 1 & 1 & 0 & 1 \\
      &   &   &   &   & 1 & 1 & 0 \\
      &   &   &   &   &   & 1 & 1 \\
      &   &   &   &   &   &   & 1 \\
};
\draw[red, thick] 
        (M1-1-1.north west) -| (M1-2-2.south east) -| (M1-1-1.north west);
\draw[red, thick] 
        (M1-3-3.north west) -| (M1-4-4.south east) -| (M1-3-3.north west);
\draw[red, thick] 
        (M1-5-5.north west) -| (M1-6-6.south east) -| (M1-5-5.north west);
\draw[red, thick] 
        (M1-7-7.north west) -| (M1-8-8.south east) -| (M1-7-7.north west);
\end{tikzpicture} 
\to 
\begin{tikzpicture}[
    node distance=1mm and 0mm,
    baseline]
 \matrix (M2) [matrix of nodes,{left delimiter=[},{right delimiter=]}]
{
    1 & 0 & 1 & 1 & 1 & 0 & 1 & 1 \\ 
      & 1 & 1 & 0 & 0 & 1 & 0 & 1 \\ 
      &   & 1 & 1 & 1 & 0 & 1 & 1 \\ 
      &   &   & 1 & 1 & 1 & 1 & 1 \\ 
      &   &   &   & 1 & 1 & 0 & 1 \\
      &   &   &   &   & 1 & 1 & 0 \\
      &   &   &   &   &   & 1 & 1 \\
      &   &   &   &   &   &   & 1 \\
};
\draw[red, thick] 
        (M2-1-1.north west) -| (M2-4-4.south east) -| (M2-1-1.north west);
\draw[red, thick] 
        (M2-5-5.north west) -| (M2-8-8.south east) -| (M2-5-5.north west);
\draw[fill=red, opacity=0.2] 
        (M2-1-1.north west) -| (M2-2-2.south east) -| (M2-1-1.north west);
\draw[fill=red, opacity=0.2] 
        (M2-3-3.north west) -| (M2-4-4.south east) -| (M2-3-3.north west);
\draw[fill=red, opacity=0.2] 
        (M2-5-5.north west) -| (M2-6-6.south east) -| (M2-5-5.north west);
\draw[fill=red, opacity=0.2] 
        (M2-7-7.north west) -| (M2-8-8.south east) -| (M2-7-7.north west);
\end{tikzpicture}
\to 
\begin{tikzpicture}[
    node distance=1mm and 0mm,
    baseline]
 \matrix (M3) [matrix of nodes,{left delimiter=[},{right delimiter=]}]
{
    1 & 0 & 1 & 1 & 1 & 0 & 1 & 1 \\ 
      & 1 & 1 & 0 & 0 & 1 & 0 & 1 \\ 
      &   & 1 & 1 & 1 & 0 & 1 & 1 \\ 
      &   &   & 1 & 1 & 1 & 1 & 1 \\ 
      &   &   &   & 1 & 1 & 0 & 1 \\
      &   &   &   &   & 1 & 1 & 0 \\
      &   &   &   &   &   & 1 & 1 \\
      &   &   &   &   &   &   & 1 \\
};
\draw[red, thick] 
        (M3-1-1.north west) -| (M3-8-8.south east) -| (M3-1-1.north west);
\draw[fill=red, opacity=0.2] 
        (M3-1-1.north west) -| (M3-4-4.south east) -| (M3-1-1.north west);
\draw[fill=red, opacity=0.2] 
        (M3-5-5.north west) -| (M3-8-8.south east) -| (M3-5-5.north west);
\end{tikzpicture}
\]
\caption{\label{fig:triangles} Illustration of the implementation strategy of an upper triangular transformation. As detailed in the main text, we first consider the $2\times 2$ regions along the diagonal. For each subsequent iteration, we multiply a factor of two to the sidelengths, exponentially expanding the areas of interest per region. 
}
\end{figure*}
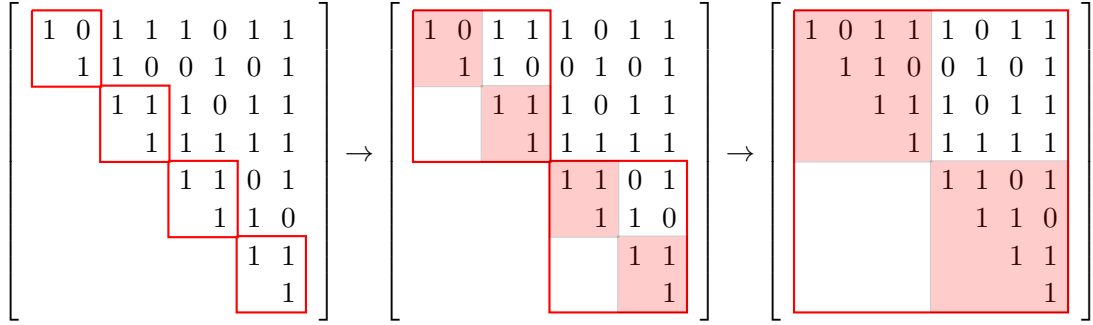

A more efficient approach is to use $n/2$ ancilla qubits, where the ancilla qubits can be used to store XORs of input Boolean variables. To start, note an upper triangular transformation takes the input Boolean variables, say $b_{i}$, $i=0,1,..,n-1$, for each qubit $i$ and outputs $o_{i} = \sum_{j=i}^{n-1} s^{(i)}_j b_j$, where $s^{(i)}_j \in \{0,1\}$ is the $j$th element of the $i$th row of the linear, triangular transformation matrix. To facilitate the foregoing discussion, we assume $n$ is a power of two, although this is not necessary.

We zoom in on the smallest triangles, regions occupying a $2\times 2$ area each, along the diagonal as shown in Fig.~\ref{fig:triangles}, of which there are $n/2$. The upper-right corner of each triangle tells us if the upper element needs to be transformed into the XOR of the upper element and the lower element of the triangle. We achieve this transformation by (a) introducing $n/2$ clean ancilla qubits initialized to $|0\rangle$, (b) computing, in this case copying, the lower element onto the ancilla, should the upper triangular element be one, (c) applying CNOTs, controlled on each ancilla, targeting each upper elements, and (d) uncomputing (b), returning all ancillae to $|0\rangle$.

We next zoom out to focus on $n/4$ triangles along the diagonal, each occupying a $4\times 4$ footprint each (see Fig.~\ref{fig:triangles}, middle panel). The idea is similar to that of the $2\times 2$ case considered above, i.e., (a)-(d). This time, each $4\times 4$ includes two $2\times 2$ triangles (shaded region) that we already took care of. Thus, all we need to consider at this point is the transformations implied by the upper-right $2 \times 2$ square (light region) of the $4\times 4$ triangle. Concretely, (a) given two clean ancillae, (b) we first compute, on the ancillae, the XOR values implied by the upper-right $2\times 2$ square using fan-ins from the lower-right $2\times 2$ triangle, of the $4 \times 4$ triangle. 
Since the qubits that correspond to the lower two elements contain XOR patterns of the elements, one can always compute the rows of the square by fan ins. Once the ancillae encode the square, (c) we impart the computed XORs onto the upper two elements by a layer of parallel CNOTs as before. Lastly, (d) we uncompute the ancillae based on the unchanged, lower two elements of each $4\times 4$ triangle.

We iterate the above process by doubling the sidelength of the triangles at each iteration. The number of ancilla qubits required remains constant at $n/2$ and the entire process is over in $\log_2(n)$ steps. Each step requires three EASE gates, one each for steps (b), (c), and (d). Since we have two triangles in the LU decomposition of a CNOT transformation layer in the Clifford normal form, we use, at most, a total of $6\log_2(n)$ EASE gates for the CNOT stage. Together with the EASE counts required for the two CZ stages, i.e., two, we use for an arbitrary Clifford transformation over $n$ qubits $6\log_2(n) + 2$ EASE gates, which may be compared to the state-of-the-art $6n$ GMS gates reported in \cite{GMS3}.

We note in passing that this construction of a CNOT stage using EASE gates can significantly benefit the near-term application of variational quantum eigensolver, simulating fermionic systems. Shown in \cite{HMP2} was the use of generalized transformation that maps fermion basis to qubit basis to optimize the quantum circuit that results in ground-state energy estimates of a given chemcial system. The optimization strategy relies on the ability to induce the transformation efficiently, i.e., the reduction in the circuit complexity due to a good choice of transformation needs to outweigh the resource cost required for implementing the transformation itself in order to gain any advantages. Note the transformation considered therein is precisely the triangular transformation considered in this section. With the EASE approach, the transformation cost is kept at minimum.

\section{Multiply-controlled NOT gates}
\label{sec:Toffoli}

Shown in \cite{GMS1} was the method to construct multiply-controlled, ${\mathrm C}^{n-1}$NOT gates (or Toffoli-$n$ gates, as used in \cite{GMS1}) using GMS gates, where for a $(n-1)$-control gate, $3n$ GMS gates were used. Later, the state of the art was superseded with the results in \cite{GMS2}, requiring only $2n$ GMS gates. In this section, we reduce the number of EASE gates required to $3n/2$. In the extreme, rather unrealistic case of nearly-free ancilla qubits and pulse design resources, we can further reduce the EASE counts to two.

We start our discussion by reiterating the point made in \cite{GMS1}, i.e., a ${\mathrm C}^{n-1}$Z gate, which is equivalent to ${\mathrm C}^{n-1}$NOT gate up to a conjugation by Hadamard gates on the target qubit, performs 
\begin{equation}
{\mathrm C}^{n-1}{\mathrm Z}|b_0,b_1,..,b_{n-1}\rangle \mapsto \left(w_{2^{n}}\right)^{2^{n-1}\prod_{j=0}^{n-1}b_j} |b_0,b_1,..,b_{n-1}\rangle,
\end{equation}
where $w_{2^{n}} := \exp(i\pi/2^{n-1})$ and we assigned qubit indices $j$ to each Boolean input $b_j$. Using $2xy = x+y-(x\oplus y)$, it was shown in \cite{GMS1} that the exponent $2^{n-1}\prod_{j=0}^{n-1}b_j$ can be expanded according to
\begin{equation}
2^{n-1}\prod_{j=0}^{n-1}b_j = \sum_{l=1}^{n} (-1)^{l-1} T_l,
\end{equation}
where
\begin{equation}
T_l = \sum_{k=1}^{n{\mathcal C}_l} \bigoplus_{m=1}^l b_{c_l(k,m)}
\end{equation}
is the sum of all distinct length-$l$ XOR patterns of input Boolean values,
$c_l(k,m)$ is the $m$'th qubit index that appears in the $k$th length-$l$ pattern, and $n{\mathcal C}_l$ is $n$ choose $l$.
Note a ZZ$(\theta)$ gate on inputs $a$ and $b$ induces, up to a global phase,
\begin{equation}
|ab\rangle \mapsto e^{i\theta (a\oplus b)} |ab\rangle.
\end{equation}
Given that an EASE gate, conjugated by a Hadamard layer, 
can implement as many ZZ gates in parallel 
as one desires with arbitrary angles for each ZZ
-- in this case $\pm \pi/2^{n-1}$ -- the problem of 
inducing a ${\mathrm C}^{n-1}$NOT gate reduces to 
inducing Boolean values such that XORs of the pairs of the values span
the entire space of $T_l$ for $l=1,2,..,n$. 

We note that, whenever possible,
we take advantage of the fact that an application of an ${\mathrm Z}^c$,
which is equivalent to $\exp(-i c \pi \sigma_z/2)$ up to a global phase,
induces
\begin{equation}
|a\rangle \mapsto e^{i \theta a} |a\rangle,
\end{equation}
where $\theta = c \pi$.
For instance, we take care of all $l=1$ terms by
simply applying appropriate ${\mathrm Z}^c$ gates
to each qubit. We use this on some of the computed
XOR patterns, to be discussed below.

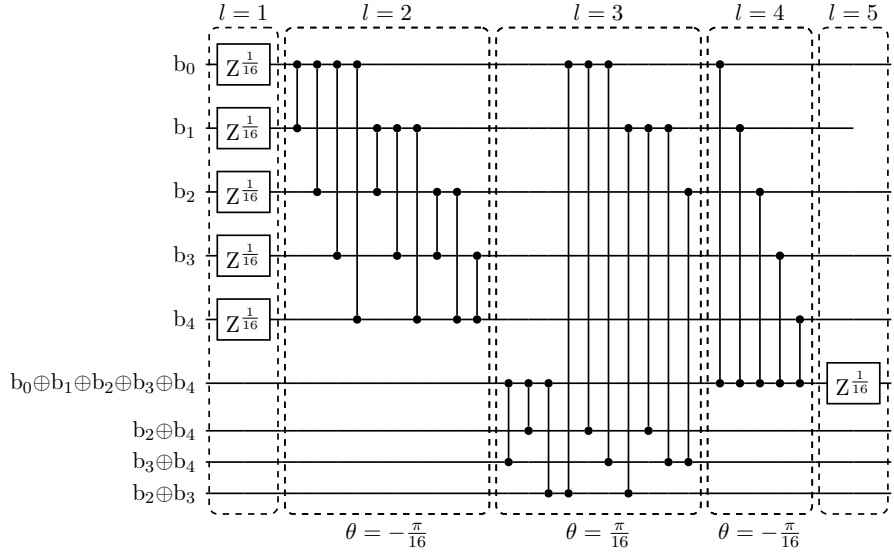
\begin{figure*}
    \centering
\vspace{-2em}
    \resizebox{0.8\textwidth}{!}{
    \begin{quantikz}[row sep=1.0em, column sep=0.5em]
\lstick{b$_0$} & \gate{\text{Z}^{\frac{1}{16}}}\gategroup[9,steps=1,style={dashed,rounded corners, inner xsep=0pt},background]{$l=1$} & \qw & \ctrl{1}\gategroup[9,steps=10,style={dashed,rounded corners, inner xsep=0pt},background]{$l=2$}\gategroup[9,steps=10,style={dashed,rounded corners, inner xsep=0pt},background,label style={label position=below,yshift=-0.2cm,anchor=north}]{$\theta=-\frac{\pi}{16}$} & \ctrl{2} & \ctrl{3} & \ctrl{4} & \qw & \qw & \qw & \qw & \qw & \qw & \qw & \qw\gategroup[9,steps=10,style={dashed,rounded corners, inner xsep=0pt},background]{$l=3$}\gategroup[9,steps=10,style={dashed,rounded corners, inner xsep=0pt},background,label style={label position=below,yshift=-0.2cm,anchor=north}]{$\theta=\frac{\pi}{16}$} & \qw & \qw & \ctrl{8} & \ctrl{6} & \ctrl{7} & \qw & \qw & \qw & \qw & \qw & \ctrl{5}\gategroup[9,steps=5,style={dashed,rounded corners, inner xsep=0pt},background]{$l=4$}\gategroup[9,steps=5,style={dashed,rounded corners, inner xsep=0pt},background,label style={label position=below,yshift=-0.2cm,anchor=north}]{$\theta=-\frac{\pi}{16}$} & \qw & \qw & \qw & \qw & \qw & \qw\gategroup[9,steps=1,style={dashed,rounded corners, inner xsep=0pt},background]{$l=5$} & \qw\\
\lstick{b$_1$} & \gate{\text{Z}^{\frac{1}{16}}} & \qw & \control{} & \qw & \qw & \qw & \ctrl{1} & \ctrl{2} & \ctrl{3} & \qw & \qw & \qw & \qw & \qw & \qw & \qw & \qw & \qw & \qw & \ctrl{7} & \ctrl{5} & \ctrl{6} & \qw & \qw & \qw & \ctrl{4} & \qw & \qw & \qw & \qw & \qw\\
\lstick{b$_2$} & \gate{\text{Z}^{\frac{1}{16}}} & \qw & \qw & \control{} & \qw & \qw & \control{} & \qw & \qw & \ctrl{1} & \ctrl{2} & \qw & \qw & \qw & \qw & \qw & \qw & \qw & \qw & \qw & \qw & \qw & \ctrl{5} & \qw & \qw & \qw & \ctrl{3} & \qw & \qw & \qw & \qw & \qw\\
\lstick{b$_3$} & \gate{\text{Z}^{\frac{1}{16}}} & \qw & \qw & \qw & \control{} & \qw & \qw & \control{} & \qw & \control{} & \qw & \ctrl{1} & \qw & \qw & \qw & \qw & \qw & \qw & \qw & \qw & \qw & \qw & \qw & \qw & \qw & \qw & \qw & \ctrl{2} & \qw & \qw & \qw & \qw\\
\lstick{b$_4$} & \gate{\text{Z}^{\frac{1}{16}}} & \qw & \qw & \qw & \qw & \control{} & \qw & \qw & \control{} & \qw & \control{} & \control{} & \qw & \qw & \qw & \qw & \qw & \qw & \qw & \qw & \qw & \qw & \qw & \qw & \qw & \qw & \qw & \qw & \ctrl{1} & \qw & \qw & \qw\\
\lstick{b$_0\oplus$b$_1\oplus$b$_2\oplus$b$_3\oplus$b$_4$} & \qw & \qw & \qw & \qw & \qw & \qw & \qw & \qw & \qw & \qw & \qw & \qw & \qw & \ctrl{2} & \ctrl{1} & \ctrl{3} & \qw & \qw & \qw & \qw & \qw & \qw & \qw & \qw & \control{} & \control{} & \control{} & \control{} & \control{} & \qw & \gate{\text{Z}^{\frac{1}{16}}} & \qw\\
\lstick{b$_2\oplus$b$_4$} & \qw & \qw & \qw & \qw & \qw & \qw & \qw & \qw & \qw & \qw & \qw & \qw & \qw & \qw & \control{} & \qw & \qw & \control{} & \qw & \qw & \control{} & \qw & \qw & \qw & \qw & \qw & \qw & \qw & \qw & \qw & \qw & \qw\\
\lstick{b$_3\oplus$b$_4$} & \qw & \qw & \qw & \qw & \qw & \qw & \qw & \qw & \qw & \qw & \qw & \qw & \qw & \control{} & \qw & \qw & \qw & \qw & \control{} & \qw & \qw & \control{} & \control{} & \qw & \qw & \qw & \qw & \qw & \qw & \qw & \qw & \qw\\
\lstick{b$_2\oplus$b$_3$} & \qw & \qw & \qw & \qw & \qw & \qw & \qw & \qw & \qw & \qw & \qw & \qw & \qw & \qw & \qw & \control{} & \control{} & \qw & \qw & \control{} & \qw & \qw & \qw & \qw & \qw & \qw & \qw & \qw & \qw & \qw & \qw & \qw
    \end{quantikz}}
    \caption{Circuit diagram relevant for an EASE-based implementation of a ${\mathrm C}^{4}$NOT gate. The circuit shown implements the phase inductions detailed in the main text. A ${\mathrm C}^{4}$NOT gate is implemented by conjugating the circuit by circuits shown in Fig.~\ref{fig:aux5} in Appendix~\ref{app:Toffoli}, then further conjugating the resulting circuit by Hadamard gates on the target qubit (unspecified). Shown are the single-qubit ${\mathrm Z}^c$ gates, which appear in solid boxes, and the ZZ gates, which appear as solid lines, each connecting two qubits marked by solid circles. Dashed boxes are drawn to delineate different pattern lengths $l$. The entanglement angles $\theta$ of ZZ gates for the different $l$ values appear at the bottom of the dashed boxes. Input Boolean variables appear to the left of the circuit.}
    \label{fig:toff5}
    \vspace{-1em}
\end{figure*}

A three-GMS construction of ${\mathrm C}^{2}$NOT and ${\mathrm C}^{3}$NOT are already available in \cite{GMS1}.
We therefore focus on ${\mathrm C}^{4}$NOT and ${\mathrm C}^{5}$NOT and show that each of them can be
constructed with three EASE gates, consuming four and seven ancillae each.
To see how the construction works, we assume we start with the input Boolean variables $b_j$'s.
For ${\mathrm C}^{4}$NOT, we induce (or compute -- used interchangeably for the Boolean variables) $b_2\oplus b_3, b_2\oplus b_4, b_3\oplus b_4$, and
$b_0\oplus b_1 \oplus b_2\oplus b_3\oplus b_4$ on to each of the four ancillae (see Fig.~\ref{fig:aux5} in Appendix~\ref{app:Toffoli}).
For ${\mathrm C}^{5}$NOT, we induce $b_0\oplus b_1\oplus b_2 \oplus b_3 \oplus b_4 \oplus b_5, b_0\oplus b_1\oplus b_2 \oplus b_5, b_0\oplus b_1\oplus b_3 \oplus b_4, b_0\oplus b_2\oplus b_3 \oplus b_4, b_1\oplus b_2\oplus b_3 \oplus b_4, b_3\oplus b_5$, and $b_4\oplus b_5$ on to each of the seven ancillae (see Fig.~\ref{fig:aux6} in Appendix~\ref{app:Toffoli}). It is then possible to show by a simple exhaustive search that an appropriately selected set of pairs of qubits can induce all of $T_l$, $l=2,3,4$ for ${\mathrm C}^{4}$NOT (Fig.~\ref{fig:toff5}). For ${\mathrm C}^{5}$NOT, the set can cover all $T_l$, $l=2,3,4,5$, except for
$b_0\oplus b_1\oplus b_2 \oplus b_5, b_0\oplus b_1\oplus b_3 \oplus b_4, b_0\oplus b_2\oplus b_3 \oplus b_4, b_1\oplus b_2\oplus b_3 \oplus b_4$, which are the XOR patterns already encoded in the ancillae (Fig.~\ref{fig:toff6}). Therefore, we take care of these four patterns by simply applying appropriate ${\mathrm Z}^c$ gates. Similarly, we take care of $T_5$ for ${\mathrm C}^{4}$NOT and $T_6$ for ${\mathrm C}^{5}$NOT using ${\mathrm Z}^c$ gates applied to the ancillae that already contain the corresponding XOR patterns.

\begin{figure*}[h!]
    \centering
    \resizebox{0.97\textwidth}{!}{
    \begin{quantikz}[row sep=1.0em, column sep=0.5em]
\lstick{b$_0$} & \gate{\text{Z}^{\frac{1}{32}}}\gategroup[6,steps=1,style={dashed,rounded corners, inner xsep=0pt},background]{$l=1$} & \qw & \ctrl{1}\gategroup[6,steps=11,style={dashed,rounded corners, inner xsep=0pt},background]{$l=2$}\gategroup[6,steps=11,style={dashed,rounded corners, inner xsep=0pt},background,label style={label position=below,yshift=-0.2cm,anchor=north}]{$\theta=-\frac{\pi}{32}$} & \ctrl{2} & \ctrl{3} & \ctrl{4} & \ctrl{5} & \qw & \qw & \qw & \qw & \qw & \qw & \qw & \qw\gategroup[13,steps=20,style={dashed,rounded corners, inner xsep=0pt},background]{$l=3$}\gategroup[13,steps=20,style={dashed,rounded corners, inner xsep=0pt},background,label style={label position=below,yshift=-0.2cm,anchor=north}]{$\theta=\frac{\pi}{32}$} & \qw & \qw & \qw & \qw & \qw & \qw & \qw & \ctrl{11} & \ctrl{12} & \qw & \qw & \ctrl{7} & \ctrl{8} & \qw & \qw & \qw & \qw & \qw & \qw & \qw & \ctrl{6}\gategroup[13,steps=6,style={dashed,rounded corners, inner xsep=0pt},background]{$l=5$}\gategroup[13,steps=6,style={dashed,rounded corners, inner xsep=0pt},background,label style={label position=below,yshift=-0.2cm,anchor=north}]{$\theta=\frac{\pi}{32}$} & \qw & \qw & \qw & \qw & \qw & \qw\\
\lstick{b$_1$} & \gate{\text{Z}^{\frac{1}{32}}} & \qw & \control{} & \qw & \qw & \qw & \qw & \ctrl{1} & \ctrl{2} & \ctrl{3} & \ctrl{4} & \qw & \qw & \qw & \qw & \qw & \qw & \qw & \qw & \qw & \ctrl{6} & \ctrl{7} & \qw & \qw & \qw & \qw & \qw & \qw & \ctrl{10} & \ctrl{11} & \ctrl{9} & \qw & \qw & \qw & \qw & \qw & \ctrl{5} & \qw & \qw & \qw & \qw & \qw\\
\lstick{b$_2$} & \gate{\text{Z}^{\frac{1}{32}}} & \qw & \ctrl{3} & \control{} & \qw & \qw & \qw & \control{} & \qw & \qw & \qw & \ctrl{1} & \ctrl{2} & \qw & \qw & \qw & \qw & \ctrl{5} & \qw & \qw & \qw & \qw & \qw & \qw & \qw & \qw & \qw & \qw & \qw & \qw & \qw & \ctrl{9} & \ctrl{10} & \qw & \qw & \qw & \qw & \ctrl{4} & \qw & \qw & \qw & \qw\\
\lstick{b$_3$} & \gate{\text{Z}^{\frac{1}{32}}} & \qw & \qw & \ctrl{1} & \control{} & \qw & \qw & \ctrl{2} & \control{} & \qw & \qw & \control{} & \qw & \qw & \qw & \qw & \ctrl{5} & \qw & \qw & \ctrl{6} & \qw & \qw & \qw & \qw & \qw & \ctrl{7} & \qw & \qw & \qw & \qw & \qw & \qw & \qw & \ctrl{9} & \qw & \qw & \qw & \qw & \ctrl{3} & \qw & \qw & \qw\\
\lstick{b$_4$} & \gate{\text{Z}^{\frac{1}{32}}} & \qw & \qw & \control{} & \qw & \control{} & \qw & \qw & \ctrl{1} & \control{} & \qw & \qw & \control{} & \qw & \qw & \ctrl{4} & \qw & \qw & \ctrl{5} & \qw & \qw & \qw & \qw & \qw & \ctrl{6} & \qw & \qw & \qw & \qw & \qw & \qw & \qw & \qw & \qw & \qw & \qw & \qw & \qw & \qw & \ctrl{2} & \qw & \qw\\
\lstick{b$_5$} & \gate{\text{Z}^{\frac{1}{32}}} & \qw & \control{} & \qw & \qw & \qw & \control{} & \control{} & \control{} & \qw & \control{} & \qw & \qw & \qw & \ctrl{2} & \qw & \qw & \qw & \qw & \qw & \qw & \qw & \qw & \qw & \qw & \qw & \qw & \qw & \qw & \qw & \qw & \qw & \qw & \qw & \qw & \qw & \qw & \qw & \qw & \qw & \ctrl{1} & \qw\\[0.75cm]
\lstick{b$_0\oplus$b$_1\oplus$b$_2\oplus$b$_3\oplus$b$_4\oplus$b$_5$} & \gate{\text{Z}^{-\frac{1}{32}}}\gategroup[1,steps=1,style={dashed,rounded corners, inner xsep=0pt},background]{$l=6$} & \qw & \qw & \qw & \qw & \qw & \qw & \qw & \qw & \qw & \qw & \qw & \qw & \qw & \qw & \qw & \qw & \qw & \qw & \qw & \qw & \qw & \qw & \qw & \qw & \qw & \qw & \qw & \qw & \qw & \qw & \qw & \qw & \qw & \qw & \control{} & \control{} & \control{} & \control{} & \control{} & \control{} & \qw\\[0.3cm]
\lstick{b$_0\oplus$b$_1\oplus$b$_2\oplus$b$_5$} & \gate{\text{Z}^{-\frac{1}{32}}}\gategroup[6,steps=13,style={dashed,rounded corners, inner xsep=0pt},background,label style={xshift=0.75cm}]{$l=4$}\gategroup[6,steps=13,style={dashed,rounded corners, inner xsep=0pt},background,label style={label position=below,yshift=-0.2cm,anchor=north,xshift=0.75cm}]{$\theta=-\frac{\pi}{32}$} & \qw & \ctrl{4} & \ctrl{5} & \qw & \qw & \qw & \qw & \ctrl{3} & \qw & \qw & \ctrl{2} & \ctrl{1} & \qw & \control{} & \qw & \qw & \control{} & \qw & \qw & \control{} & \qw & \qw & \qw & \qw & \qw & \control{} & \qw & \qw & \qw & \qw & \qw & \qw & \qw & \qw & \qw & \qw & \qw & \qw & \qw & \qw & \qw\\
\lstick{b$_0\oplus$b$_1\oplus$b$_3\oplus$b$_4$} & \gate{\text{Z}^{-\frac{1}{32}}} & \qw & \qw & \qw & \ctrl{4} & \ctrl{3} & \qw & \qw & \qw & \qw & \qw & \qw & \control{} & \qw & \qw & \control{} & \control{} & \qw & \qw & \qw & \qw & \control{} & \qw & \qw & \qw & \qw & \qw & \control{} & \qw & \qw & \qw & \qw & \qw & \qw & \qw & \qw & \qw & \qw & \qw & \qw & \qw & \qw\\
\lstick{b$_0\oplus$b$_2\oplus$b$_3\oplus$b$_4$} & \gate{\text{Z}^{-\frac{1}{32}}} & \qw & \qw & \qw & \qw & \qw & \ctrl{3} & \ctrl{2} & \qw & \qw & \qw & \control{} & \qw & \qw & \qw & \qw & \qw & \qw & \control{} & \control{} & \qw & \qw & \qw & \qw & \qw & \qw & \qw & \qw & \qw & \qw & \qw & \qw & \qw & \qw & \qw & \qw & \qw & \qw & \qw & \qw & \qw & \qw\\
\lstick{b$_1\oplus$b$_2\oplus$b$_3\oplus$b$_4$} & \gate{\text{Z}^{-\frac{1}{32}}} & \qw & \qw & \qw & \qw & \qw & \qw & \qw & \control{} & \ctrl{2} & \ctrl{1} & \qw & \qw & \qw & \qw & \qw & \qw & \qw & \qw & \qw & \qw & \qw & \qw & \qw & \control{} & \control{} & \qw & \qw & \qw & \qw & \control{} & \qw & \qw & \qw & \qw & \qw & \qw & \qw & \qw & \qw & \qw & \qw\\
\lstick{b$_3\oplus$b$_5$} & \qw & \qw & \control{} & \qw & \qw & \control{} & \qw & \control{} & \qw & \qw & \control{} & \qw & \qw & \qw & \qw & \qw & \qw & \qw & \qw & \qw & \qw & \qw & \control{} & \qw & \qw & \qw & \qw & \qw & \control{} & \qw & \qw & \control{} & \qw & \qw & \qw & \qw & \qw & \qw & \qw & \qw & \qw & \qw\\
\lstick{b$_4\oplus$b$_5$} & \qw & \qw & \qw & \control{} & \control{} & \qw & \control{} & \qw & \qw & \control{} & \qw & \qw & \qw & \qw & \qw & \qw & \qw & \qw & \qw & \qw & \qw & \qw & \qw & \control{} & \qw & \qw & \qw & \qw & \qw & \control{} & \qw & \qw & \control{} & \control{} & \qw & \qw & \qw & \qw & \qw & \qw & \qw & \qw
    \end{quantikz}}
    \caption{Circuit diagram relevant for an EASE-based implementation of a ${\mathrm C}^{5}$NOT gate. The circuit shown implements the phase inductions detailed in the main text. A ${\mathrm C}^{5}$NOT gate is implemented by conjugating the circuit by circuits shown in Fig.~\ref{fig:aux6} in Appendix~\ref{app:Toffoli}, then further conjugating the resulting circuit by Hadamard gates on the target qubit (unspecified). The description of the figure is the same as in Fig.~\ref{fig:toff5}. For the convenience of the readers we import the caption verbatim. The circuit implements the phase inductions detailed in the main text. Shown are the single-qubit ${\mathrm Z}^c$ gates, which appear in solid boxes, and the ZZ gates, which appear as solid lines, each connecting two qubits marked by solid circles. Dashed boxes are drawn to delineate different pattern lengths $l$. The entanglement angles $\theta$ of ZZ gates for the different $l$ values appear at the bottom of the dashed boxes. Input Boolean variables appear to the left of the circuit.}
    \label{fig:toff6}
\end{figure*}

Since the induction of the appropriate XOR patterns onto ancillae cost a single EASE gate, as they are simply a parallel fan-in operations onto each ancilla qubits, computing and uncomputing the patterns cost two EASE gates. Combined with the parallel ZZ operation described above, which costs a single EASE gate, we arrive at a three-EASE construction of ${\mathrm C}^{4}$NOT and ${\mathrm C}^{5}$NOT.

By use of our ${\mathrm C}^{4}$NOT and ${\mathrm C}^{5}$NOT gates, one can employ a similar approach explored in \cite{GMS1} to extend the number of controls to an arbitrary number. Note a ${\mathrm C}^{n-1}$NOT essentially computes the AND of the $n-1$ controls. This can thus be implemented by the use of multiple ${\mathrm C}^{4}$NOT and ${\mathrm C}^{5}$NOT gates, e.g., by computing and aggregating the AND values of the subsets of size four or five of the controls. Following the method in \cite{GMS1}, we can see that, by use of ${\mathrm C}^{5}$NOT gates, roughly four controls can be taken care of each time it is used. Provided that a nested ${\mathrm C}^{5}$NOT gates need to be uncomputed, we find $n/2$ ${\mathrm C}^{5}$NOT gates are needed for a ${\mathrm C}^{n-1}$NOT. Since in our construction each ${\mathrm C}^{5}$NOT requires three EASE gates, overall we require $3n/2$ EASE gates per ${\mathrm C}^{n-1}$NOT, with the ancilla count being $n/4$ up to a small additive constant, no more than seven.

In the extreme (unrealistic) scenario where ancilla qubits are inexpensive and the pulse design with exponential complexity is not an issue, a ${\mathrm C}^{n-1}$NOT gate, regardless of $n$, can straightforwardly be implemented using only two EASE gates. To see this, we introduce $2^n-n-1$ ancilla qubits. For each ancilla qubit we compute $\bigoplus_{m=1}^l b_{c_l(k,m)}$, for all $l>1$ and $k$, of which there are $2^n-n-1$. We can then induce appropriate phase to each XOR pattern by a layer of ${\mathrm Z}^c$ gates. We uncompute and free the ancilla qubits. The number of EASE gates required is two, since the compute-uncompute operations require parallel fan-in operations, all controls being the original input, each target being ancilla.

We note in passing that the two-EASE based multiply-controlled NOT gate, like its three-EASE based counterpart, can also be used to extend the number of controls to an arbitrary number. Consider a ${\mathrm C}^{b-1}$NOT gate, requiring two EASE gates and $2^b-b-1$ ancilla qubits. A ${\mathrm C}^{n-1}$NOT gate, built with a plurality of the two-EASE based ${\mathrm C}^{b-1}$NOT gates, $n \gg b$, would then require, approximately, $4n/(b-2)$ EASE gates and $n/(b-2) + 2^b-b-1$ ancilla qubits. For $b=6$, the case we considered earlier, they amount to $n$ EASE gates and $n/4+57$ ancilla qubits. The latter number indeed motivates our three-EASE based construction, keeping the ancilla counts modest.

\section{Qubit Permutations}
\label{sec:permutation}

Another quantum subroutine where EASE gates offer an improvement over the known state-of-start is the implementation of permutations over qubit registers. It is known, from \cite{Permutation}, that any permutation over $n$ qubits can be implemented as four layers of disjoint transpositions with $n$ ancillae and, alternatively, as six layers of disjoint transpositions with no ancillae. Since each transposition, or a SWAP gate, can be implemented as three CNOT gates, each layer of disjoint transpositions can be implemented as three layers of EASE gates. Therefore, any permutation over $n$ qubits can be implemented with $O(1)$ EASE gates. 

We next consider a controlled permutation operation. Note that, since the $O(1)$ transpositions used to apply a permutation are disjoint, the $O(1)$ controlled transpositions that comprise a controlled permutation operation are disjoint as well, up to the shared control. To achieve the EASE-based implementation, 
note that a controlled-transposition gate can be implemented using seven CNOT gates \cite{niroula2021quantum}. The CNOT gates, forming seven stages, then comprise a single layer of controlled-transposition gates, together with single qubit gates. The CNOT gates in each CNOT stage are either disjoint (conventional parallel CNOTs) or they collide at the shared control (a fan-out operation). Both of these can be implemented in parallel using a single EASE gate. This implies that a controlled permutation can be implemented using $O(1)$ EASE gates. 

Using this technique, we can reduce the complexity of algorithms that use controlled permutations. As an example, we consider the circuit depth of string-matching algorithm \cite{niroula2021quantum}, which matches a pattern of length $M$ in a text of length $N$. Our EASE-based approach can reduce the depth from the reported $O\left(\sqrt{N}((\log N)^2 + \log M)\right)$ to $O\left(\sqrt{N}(\log N + \log M)\right)$. The number of ancilla qubits is reduced by $\log N$ as well. 

\section{Outlook}

Harking back to the power and impact of parallel operations
in conventional computing, we look forward to the
revolution that EASE gates are to bring in quantum computing.
Experimental demonstrations have already been
successful~\cite{EASE}, and the exciting challenges
of finding more use cases of EASE gates continue. 
In this paper, we took a first step
towards this goal, leveraging in particular
the random access over any pairs of qubits.
A future endeavor could include
more sophisticated uses of real degrees of freedom in both the amount of
entanglement induced between the pairs and the axis of rotation for
each Pauli operator acting on each participating qubit in the EASE gates.
We hope our work serves as a stepping stone
towards efficient quantum computing,
adding to the old adage of ``circuit depth matters.''

\section*{Data availability}
All data needed to evaluate the conclusions in the paper 
are present in the paper and/or the Supplementary Information. 
Additional data related to this paper may be requested from the authors. Correspondence and requests for materials should be addressed to Y.N.(ynam@umd.edu).

\section*{Acknowledgements}
We thank Qingfeng Wang 
at the University of Maryland,
Christopher Monroe at Duke University,
and Igor Markov at the University of
Michigan for helpful discussions.

\section*{Author contribution}
N.G., A.M., and P.N. contributed to devising the methods to optimize various quantum circuit constructions using EASE gates under Y.N.'s supervision. All authors contributed to preparing the manuscript. The authors declare that they have no competing interests.

\bibliography{citations}

\begin{thebibliography}{10}%
\makeatletter
\providecommand \@ifxundefined [1]{%
 \@ifx{#1\undefined}
}%
\providecommand \@ifnum [1]{%
 \ifnum #1\expandafter \@firstoftwo
 \else \expandafter \@secondoftwo
 \fi
}%
\providecommand \@ifx [1]{%
 \ifx #1\expandafter \@firstoftwo
 \else \expandafter \@secondoftwo
 \fi
}%
\providecommand \natexlab [1]{#1}%
\providecommand \enquote  [1]{``#1''}%
\providecommand \bibnamefont  [1]{#1}%
\providecommand \bibfnamefont [1]{#1}%
\providecommand \citenamefont [1]{#1}%
\providecommand \href@noop [0]{\@secondoftwo}%
\providecommand \href [0]{\begingroup \@sanitize@url \@href}%
\providecommand \@href[1]{\@@startlink{#1}\@@href}%
\providecommand \@@href[1]{\endgroup#1\@@endlink}%
\providecommand \@sanitize@url [0]{\catcode `\\12\catcode `\$12\catcode
  `\&12\catcode `\#12\catcode `\^12\catcode `\_12\catcode `\%12\relax}%
\providecommand \@@startlink[1]{}%
\providecommand \@@endlink[0]{}%
\providecommand \url  [0]{\begingroup\@sanitize@url \@url }%
\providecommand \@url [1]{\endgroup\@href {#1}{\urlprefix }}%
\providecommand \urlprefix  [0]{URL }%
\providecommand \Eprint [0]{\href }%
\providecommand \doibase [0]{http://dx.doi.org/}%
\providecommand \selectlanguage [0]{\@gobble}%
\providecommand \bibinfo  [0]{\@secondoftwo}%
\providecommand \bibfield  [0]{\@secondoftwo}%
\providecommand \translation [1]{[#1]}%
\providecommand \BibitemOpen [0]{}%
\providecommand \bibitemStop [0]{}%
\providecommand \bibitemNoStop [0]{.\EOS\space}%
\providecommand \EOS [0]{\spacefactor3000\relax}%
\providecommand \BibitemShut  [1]{\csname bibitem#1\endcsname}%
\let\auto@bib@innerbib\@empty
\bibitem [{\citenamefont {Hughes}(2015)}]{hughes2015single}%
  \BibitemOpen
  \bibfield  {author} {\bibinfo {author} {\bibfnamefont {C.~J.}\ \bibnamefont
  {Hughes}},\ }\href {\doibase 10.2200/S00647ED1V01Y201505CAC032} {\bibfield
  {journal} {\bibinfo  {journal} {Synthesis Lectures on Computer Architecture}\
  }\textbf {\bibinfo {volume} {10}},\ \bibinfo {pages} {1} (\bibinfo {year}
  {2015})}\BibitemShut {NoStop}%
\bibitem [{\citenamefont {Grzesiak}\ \emph {et~al.}(2020)\citenamefont
  {Grzesiak}, \citenamefont {Bl{\"u}mel}, \citenamefont {Wright}, \citenamefont
  {Beck}, \citenamefont {Pisenti}, \citenamefont {Li}, \citenamefont {Chaplin},
  \citenamefont {Amini}, \citenamefont {Debnath}, \citenamefont {Chen} \emph
  {et~al.}}]{EASE}%
  \BibitemOpen
  \bibfield  {author} {\bibinfo {author} {\bibfnamefont {N.}~\bibnamefont
  {Grzesiak}}, \bibinfo {author} {\bibfnamefont {R.}~\bibnamefont
  {Bl{\"u}mel}}, \bibinfo {author} {\bibfnamefont {K.}~\bibnamefont {Wright}},
  \bibinfo {author} {\bibfnamefont {K.~M.}\ \bibnamefont {Beck}}, \bibinfo
  {author} {\bibfnamefont {N.~C.}\ \bibnamefont {Pisenti}}, \bibinfo {author}
  {\bibfnamefont {M.}~\bibnamefont {Li}}, \bibinfo {author} {\bibfnamefont
  {V.}~\bibnamefont {Chaplin}}, \bibinfo {author} {\bibfnamefont {J.~M.}\
  \bibnamefont {Amini}}, \bibinfo {author} {\bibfnamefont {S.}~\bibnamefont
  {Debnath}}, \bibinfo {author} {\bibfnamefont {J.-S.}\ \bibnamefont {Chen}},
  \emph {et~al.},\ }\href {\doibase 10.1038/s41467-020-16790-9} {\bibfield
  {journal} {\bibinfo  {journal} {Nature communications}\ }\textbf {\bibinfo
  {volume} {11}},\ \bibinfo {pages} {1} (\bibinfo {year} {2020})}\BibitemShut
  {NoStop}%
\bibitem [{\citenamefont {Maslov}\ and\ \citenamefont {Nam}(2018)}]{GMS1}%
  \BibitemOpen
  \bibfield  {author} {\bibinfo {author} {\bibfnamefont {D.}~\bibnamefont
  {Maslov}}\ and\ \bibinfo {author} {\bibfnamefont {Y.}~\bibnamefont {Nam}},\
  }\href {\doibase 10.1088/1367-2630/aaa398} {\bibfield  {journal} {\bibinfo
  {journal} {New Journal of Physics}\ }\textbf {\bibinfo {volume} {20}},\
  \bibinfo {pages} {033018} (\bibinfo {year} {2018})}\BibitemShut {NoStop}%
\bibitem [{\citenamefont {Groenland}\ \emph {et~al.}(2020)\citenamefont
  {Groenland}, \citenamefont {Witteveen}, \citenamefont {Schoutens},\ and\
  \citenamefont {Gerritsma}}]{GMS2}%
  \BibitemOpen
  \bibfield  {author} {\bibinfo {author} {\bibfnamefont {K.}~\bibnamefont
  {Groenland}}, \bibinfo {author} {\bibfnamefont {F.}~\bibnamefont
  {Witteveen}}, \bibinfo {author} {\bibfnamefont {K.}~\bibnamefont
  {Schoutens}}, \ and\ \bibinfo {author} {\bibfnamefont {R.}~\bibnamefont
  {Gerritsma}},\ }\href {\doibase 10.1088/1367-2630/ab8830} {\bibfield
  {journal} {\bibinfo  {journal} {New Journal of Physics}\ }\textbf {\bibinfo
  {volume} {22}},\ \bibinfo {pages} {063006} (\bibinfo {year}
  {2020})}\BibitemShut {NoStop}%
\bibitem [{\citenamefont {van~de Wetering}(2021)}]{GMS3}%
  \BibitemOpen
  \bibfield  {author} {\bibinfo {author} {\bibfnamefont {J.}~\bibnamefont
  {van~de Wetering}},\ }\href {\doibase 10.1088/1367-2630/abf1b3} {\bibfield
  {journal} {\bibinfo  {journal} {New Journal of Physics}\ }\textbf {\bibinfo
  {volume} {23}},\ \bibinfo {pages} {043015} (\bibinfo {year}
  {2021})}\BibitemShut {NoStop}%
\bibitem [{\citenamefont {Zhu}\ \emph {et~al.}(2006)\citenamefont {Zhu},
  \citenamefont {Monroe},\ and\ \citenamefont {Duan}}]{Zhu_2006}%
  \BibitemOpen
  \bibfield  {author} {\bibinfo {author} {\bibfnamefont {S.-L.}\ \bibnamefont
  {Zhu}}, \bibinfo {author} {\bibfnamefont {C.}~\bibnamefont {Monroe}}, \ and\
  \bibinfo {author} {\bibfnamefont {L.-M.}\ \bibnamefont {Duan}},\ }\href
  {\doibase 10.1209/epl/i2005-10424-4} {\bibfield  {journal} {\bibinfo
  {journal} {Europhysics Letters (EPL)}\ }\textbf {\bibinfo {volume} {73}},\
  \bibinfo {pages} {485–491} (\bibinfo {year} {2006})}\BibitemShut {NoStop}%
\bibitem [{\citenamefont {Duncan}\ \emph {et~al.}(2020)\citenamefont {Duncan},
  \citenamefont {Kissinger}, \citenamefont {Perdrix},\ and\ \citenamefont {Van
  De~Wetering}}]{CliffordNormalForm}%
  \BibitemOpen
  \bibfield  {author} {\bibinfo {author} {\bibfnamefont {R.}~\bibnamefont
  {Duncan}}, \bibinfo {author} {\bibfnamefont {A.}~\bibnamefont {Kissinger}},
  \bibinfo {author} {\bibfnamefont {S.}~\bibnamefont {Perdrix}}, \ and\
  \bibinfo {author} {\bibfnamefont {J.}~\bibnamefont {Van De~Wetering}},\
  }\href {\doibase 10.22331/q-2020-06-04-279} {\bibfield  {journal} {\bibinfo
  {journal} {Quantum}\ }\textbf {\bibinfo {volume} {4}},\ \bibinfo {pages}
  {279} (\bibinfo {year} {2020})}\BibitemShut {NoStop}%
\bibitem [{\citenamefont {Wang}\ \emph {et~al.}(2021)\citenamefont {Wang},
  \citenamefont {Li}, \citenamefont {Monroe},\ and\ \citenamefont
  {Nam}}]{HMP2}%
  \BibitemOpen
  \bibfield  {author} {\bibinfo {author} {\bibfnamefont {Q.}~\bibnamefont
  {Wang}}, \bibinfo {author} {\bibfnamefont {M.}~\bibnamefont {Li}}, \bibinfo
  {author} {\bibfnamefont {C.}~\bibnamefont {Monroe}}, \ and\ \bibinfo {author}
  {\bibfnamefont {Y.}~\bibnamefont {Nam}},\ }\href {\doibase
  10.22331/q-2021-07-26-509} {\bibfield  {journal} {\bibinfo  {journal}
  {Quantum}\ }\textbf {\bibinfo {volume} {5}},\ \bibinfo {pages} {509}
  (\bibinfo {year} {2021})}\BibitemShut {NoStop}%
\bibitem [{\citenamefont {Moore}\ and\ \citenamefont
  {Nilsson}(2001)}]{Permutation}%
  \BibitemOpen
  \bibfield  {author} {\bibinfo {author} {\bibfnamefont {C.}~\bibnamefont
  {Moore}}\ and\ \bibinfo {author} {\bibfnamefont {M.}~\bibnamefont
  {Nilsson}},\ }\href {\doibase 10.1137/S0097539799355053} {\bibfield
  {journal} {\bibinfo  {journal} {SIAM Journal on Computing}\ }\textbf
  {\bibinfo {volume} {31}},\ \bibinfo {pages} {799} (\bibinfo {year}
  {2001})}\BibitemShut {NoStop}%
\bibitem [{\citenamefont {Niroula}\ and\ \citenamefont
  {Nam}(2021)}]{niroula2021quantum}%
  \BibitemOpen
  \bibfield  {author} {\bibinfo {author} {\bibfnamefont {P.}~\bibnamefont
  {Niroula}}\ and\ \bibinfo {author} {\bibfnamefont {Y.}~\bibnamefont {Nam}},\
  }\href {\doibase 10.1038/s41534-021-00369-3} {\bibfield  {journal} {\bibinfo
  {journal} {npj Quantum Information}\ }\textbf {\bibinfo {volume} {7}},\
  \bibinfo {pages} {1} (\bibinfo {year} {2021})}\BibitemShut {NoStop}%
\end{thebibliography}%

\appendix

\section{CNOT networks for multiply-controlled NOT gates}
\label{app:Toffoli}

\begin{figure*}[h!]
    \centering
    \resizebox{\textwidth}{!}{
    \begin{quantikz}[row sep=1.0em, column sep=0.5em]
\lstick{b$_0$} & \ctrl{6} & \qw & \qw & \qw & \qw & \qw & \ctrl{7} & \qw & \qw & \qw & \ctrl{8} & \qw & \qw & \qw & \ctrl{9} & \qw & \qw & \qw & \qw & \qw & \qw & \qw & \qw & \qw & \qw & \qw & \qw\rstick{b$_0$}\\
\lstick{b$_1$} & \qw & \ctrl{5} & \qw & \qw & \qw & \qw & \qw & \ctrl{6} & \qw & \qw & \qw & \ctrl{7} & \qw & \qw & \qw & \qw & \qw & \qw & \ctrl{9} & \qw & \qw & \qw & \qw & \qw & \qw & \qw & \qw\rstick{b$_1$}\\
\lstick{b$_2$} & \qw & \qw & \ctrl{4} & \qw & \qw & \qw & \qw & \qw & \ctrl{5} & \qw & \qw & \qw & \qw & \qw & \qw & \ctrl{7} & \qw & \qw & \qw & \ctrl{8} & \qw & \qw & \qw & \qw & \qw & \qw & \qw\rstick{b$_2$}\\
\lstick{b$_3$} & \qw & \qw & \qw & \ctrl{3} & \qw & \qw & \qw & \qw & \qw & \qw & \qw & \qw & \ctrl{5} & \qw & \qw & \qw & \ctrl{6} & \qw & \qw & \qw & \ctrl{7} & \qw & \ctrl{8} & \qw & \qw & \qw & \qw\rstick{b$_3$}\\
\lstick{b$_4$} & \qw & \qw & \qw & \qw & \ctrl{2} & \qw & \qw & \qw & \qw & \qw & \qw & \qw & \qw & \ctrl{4} & \qw & \qw & \qw & \ctrl{5} & \qw & \qw & \qw & \ctrl{6} & \qw & \qw & \ctrl{8} & \qw & \qw\rstick{b$_4$}\\
\lstick{b$_5$} & \qw & \qw & \qw & \qw & \qw & \ctrl{1} & \qw & \qw & \qw & \ctrl{2} & \qw & \qw & \qw & \qw & \qw & \qw & \qw & \qw & \qw & \qw & \qw & \qw & \qw & \ctrl{6} & \qw & \ctrl{7} & \qw\rstick{b$_5$}\\
\lstick{$\ket{0}$} & \targ{} & \targ{} & \targ{} & \targ{} & \targ{} & \targ{} & \qw & \qw & \qw & \qw & \qw & \qw & \qw & \qw & \qw & \qw & \qw & \qw & \qw & \qw & \qw & \qw & \qw & \qw & \qw & \qw & \qw\rstick{b$_0\oplus$b$_1\oplus$b$_2\oplus$b$_3\oplus$b$_4\oplus$b$_5$}\\
\lstick{$\ket{0}$} & \qw & \qw & \qw & \qw & \qw & \qw & \targ{} & \targ{} & \targ{} & \targ{} & \qw & \qw & \qw & \qw & \qw & \qw & \qw & \qw & \qw & \qw & \qw & \qw & \qw & \qw & \qw & \qw & \qw\rstick{b$_0\oplus$b$_1\oplus$b$_2\oplus$b$_5$}\\
\lstick{$\ket{0}$} & \qw & \qw & \qw & \qw & \qw & \qw & \qw & \qw & \qw & \qw & \targ{} & \targ{} & \targ{} & \targ{} & \qw & \qw & \qw & \qw & \qw & \qw & \qw & \qw & \qw & \qw & \qw & \qw & \qw\rstick{b$_0\oplus$b$_1\oplus$b$_3\oplus$b$_4$}\\
\lstick{$\ket{0}$} & \qw & \qw & \qw & \qw & \qw & \qw & \qw & \qw & \qw & \qw & \qw & \qw & \qw & \qw & \targ{} & \targ{} & \targ{} & \targ{} & \qw & \qw & \qw & \qw & \qw & \qw & \qw & \qw & \qw\rstick{b$_0\oplus$b$_2\oplus$b$_3\oplus$b$_4$}\\
\lstick{$\ket{0}$} & \qw & \qw & \qw & \qw & \qw & \qw & \qw & \qw & \qw & \qw & \qw & \qw & \qw & \qw & \qw & \qw & \qw & \qw & \targ{} & \targ{} & \targ{} & \targ{} & \qw & \qw & \qw & \qw & \qw\rstick{b$_1\oplus$b$_2\oplus$b$_3\oplus$b$_4$}\\
\lstick{$\ket{0}$} & \qw & \qw & \qw & \qw & \qw & \qw & \qw & \qw & \qw & \qw & \qw & \qw & \qw & \qw & \qw & \qw & \qw & \qw & \qw & \qw & \qw & \qw & \targ{} & \targ{} & \qw & \qw & \qw\rstick{b$_3\oplus$b$_5$}\\
\lstick{$\ket{0}$} & \qw & \qw & \qw & \qw & \qw & \qw & \qw & \qw & \qw & \qw & \qw & \qw & \qw & \qw & \qw & \qw & \qw & \qw & \qw & \qw & \qw & \qw & \qw & \qw & \targ{} & \targ{} & \qw\rstick{b$_4\oplus$b$_5$}
    \end{quantikz}}
    \caption{Circuit diagram for computing (uncomputing) ancilla qubit states relevant for our ${\mathrm C}^{5}$NOT construction.}
    \label{fig:aux6}
\end{figure*}
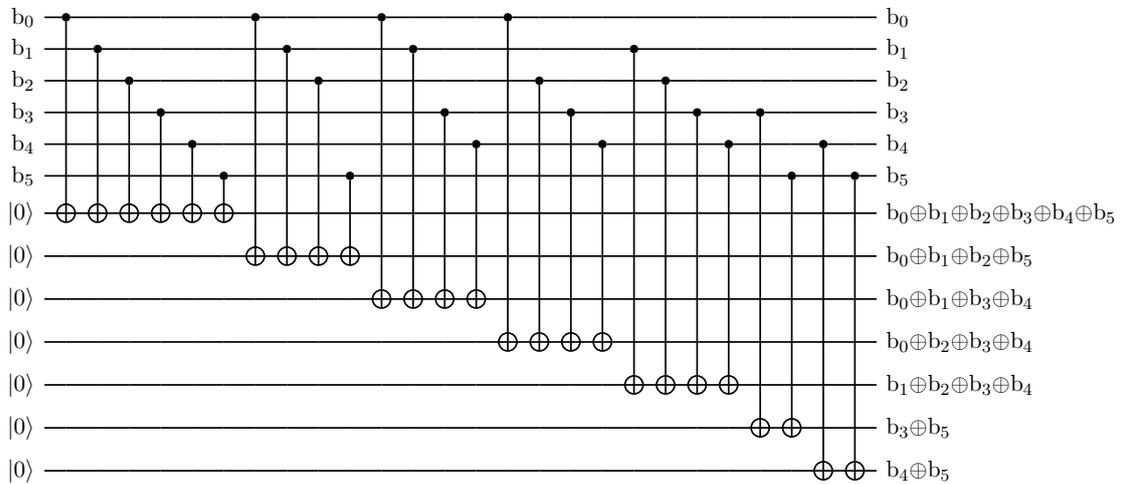

In this section, we show CNOT networks used to induce
the input Boolean variables we use for our EASE-based
implementation of multiply-controlled NOT gates.
Figure~\ref{fig:aux5} shows the network for a ${\mathrm C}^4$NOT gate.
Figure~\ref{fig:aux6} shows the network for a ${\mathrm C}^5$NOT gate.
The two networks are used before and after the phase induction
steps discussed in detail in the main text to implement 
multiply-controlled Z gates of respective sizes, while freeing
up the ancillae. 
The resulting circuits are conjugated by Hadamard gates
on the target qubit of choice in order to induce
multiply-controlled NOT gates.

\begin{figure*}[h!]
    \centering
    \begin{quantikz}[row sep=1.0em, column sep=0.5em]
\lstick{b$_0$} & \ctrl{5} & \qw & \qw & \qw & \qw & \qw & \qw & \qw & \qw & \qw & \qw & \qw\rstick{b$_0$}\\
\lstick{b$_1$} & \qw & \ctrl{4} & \qw & \qw & \qw & \qw & \qw & \qw & \qw & \qw & \qw & \qw\rstick{b$_1$}\\
\lstick{b$_2$} & \qw & \qw & \ctrl{3} & \qw & \qw & \ctrl{4} & \qw & \qw & \qw & \ctrl{6} & \qw & \qw\rstick{b$_2$}\\
\lstick{b$_3$} & \qw & \qw & \qw & \ctrl{2} & \qw & \qw & \qw & \ctrl{4} & \qw & \qw & \ctrl{5} & \qw\rstick{b$_3$}\\
\lstick{b$_4$} & \qw & \qw & \qw & \qw & \ctrl{1} & \qw & \ctrl{2} & \qw & \ctrl{3} & \qw & \qw & \qw\rstick{b$_4$}\\
\lstick{$\ket{0}$} & \targ{} & \targ{} & \targ{} & \targ{} & \targ{} & \qw & \qw & \qw & \qw & \qw & \qw & \qw\rstick{b$_0\oplus$b$_1\oplus$b$_2\oplus$b$_3\oplus$b$_4$}\\
\lstick{$\ket{0}$} & \qw & \qw & \qw & \qw & \qw & \targ{} & \targ{} & \qw & \qw & \qw & \qw & \qw\rstick{b$_2\oplus$b$_4$}\\
\lstick{$\ket{0}$} & \qw & \qw & \qw & \qw & \qw & \qw & \qw & \targ{} & \targ{} & \qw & \qw & \qw\rstick{b$_3\oplus$b$_4$}\\
\lstick{$\ket{0}$} & \qw & \qw & \qw & \qw & \qw & \qw & \qw & \qw & \qw & \targ{} & \targ{} & \qw\rstick{b$_2\oplus$b$_3$}
    \end{quantikz}
    \caption{Circuit diagram for computing (uncomputing) ancilla qubit states relevant for our ${\mathrm C}^{4}$NOT construction.}
    \label{fig:aux5}
\end{figure*}
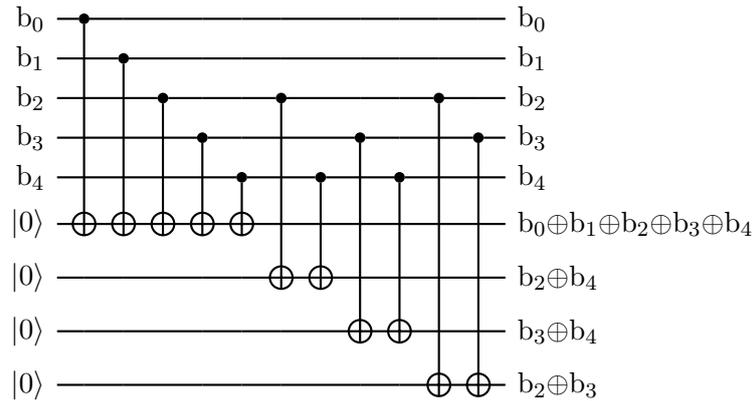

\end{document}